\documentstyle[12pt,aaspp4]{article}

\lefthead{Wiebe and Watson}
\righthead{Non-Zeeman Interpretation for Polarized Radiation}

\begin{document}

\title{A Non-Zeeman Interpretation for Polarized Maser 
Radiation\\
and the Magnetic Field at the Atmospheres of Late-Type Giants}

\author{D. S. Wiebe\altaffilmark{1} and W.D. Watson}
\affil{Department of Physics, University of Illinois at
Urbana-Champaign, Urbana, IL 61801}
\altaffiltext{1}{Permanent address: Institute of Astronomy of the RAS, 48, 
Pyatnitskaya str., 109017, Moscow, Russia}

\begin{abstract}
The linear polarization that is observed, together with likely changes
in the orientation of the magnetic field along the line of sight and hence 
of the optical axes of
the medium, can lead to the circular polarization
that is observed in the radiation of the circumstellar SiO masers. A
magnetic field greater than only about 30~mG is required, in contrast  to
10-100~G that would be implied by the Zeeman interpretation. To assess
quantitatively the likely changes in orientation of the magnetic field,
calculations are performed with representative field configurations that 
are
created by statistical sampling using a Kolmogorov-like power spectrum.
\end{abstract}


\section{Introduction}

The circular polarization of SiO maser radiation is observed to be
about 3\% when averaged over the circumstellar shell of a
representative late-type giant. It reaches a few tens of percent
in particular features of this and other late-type giants
(Barvainis, McIntosh \& Predmore 1987\markcite{BMP}; Kemball
\& Diamond 1997\markcite{KD97}, hereafter KD). In the standard
interpretation that circular
polarization is due to the Zeeman effect, the magnetic fields
would be 10-100~G and would cause a pressure that greatly exceeds
the expected thermal pressure in the masing gas. The averaged linear
polarization is about 30\%. In ordinary optics,
circularly polarized radiation is created when the electric field of 
linearly polarized
radiation is not along one of the optical axes of the medium in which it 
is propagating (e.g., Rossi 1957\markcite{R}). In astronomy, circular 
polarization in starlight is created in this way as a result of the 
changing orientation of the alignment
of dust grains caused by changing directions of the galactic magnetic 
field (e.g., Spitzer 1978\markcite{SP}). Near the convective  atmospheres
of late-type giants, it seems especially likely that irregularities
will occur in the magnetic field structure. To have distinct optical
axes, the medium must be anisotropic. The observation of the strong
linear polarization of the SiO maser radiation indicates that the
medium is anisotropic at this frequency. This anisotropy and hence
the linear polarization is almost certainly the result of
anisotropic pumping of the magnetic substates caused by an
anisotropic angular distribution of the infrared radiation that is
involved in determining the populations of the masing
states (Western \& Watson 1983, 1984\markcite{ww83,84}; Nedoluha \& Watson 
1990\markcite{NW90}).

In a previous investigation, we (Nedoluha \& Watson 1994\markcite{NW94}) 
have
shown that the rotation of the optical axes necessary for creating
the observed circular polarization can be caused by the
competition between the magnetic field $B$ and the beam of maser
radiation. This occurs when the Zeeman splitting parameter
$g\Omega$ in frequency units [$g\Omega=1.5B$(mG)s$^{-1}$ for these SiO 
masers]
is comparable with the rate $R$ for stimulated emission by the
maser radiation, and $R$ is greater than or equal to the decay rate $\Gamma$
for the molecular states so that the maser is at least partially saturated
($\Gamma=5$ s$^{-1}$ is commonly accepted for these SiO masers).
Further exploration of the cause for
the circular polarization nevertheless seems to be warranted. Even
for modest $B$, $g\Omega\gg R$ is likely instead of the required
$g\Omega\simeq R$. In addition, when $g\Omega\simeq R$ calculational 
methods
(involving density matrices) are necessary that are unfamiliar in
astrophysics. As a result, the ``non-Zeeman possiblity'' seems to
have received less attention than it merits. When
$g\Omega\gg R$, the physics is straightforward because the
relevant equations
for the transport of polarized maser radiation are identical with those 
for ordinary radiation
except for the sign of the optical depth. Circularly polarized  radiation
is then created from linear in a manner that is essentially
identical with the familiar results from ordinary optics for
likely changes in the orientation of the magnetic field along the
line of sight. 

\section{Calculations}

The relevant, coupled differential equations of radiative transport for
the Stokes intensities to describe polarized radiation, including the
possibility of a changing magnetic field and anisotropic populations, are
given in convenient form by Deguchi \& Watson (1985, esp. Section
IIIc\markcite{dw85}) and by Wallin \& Watson (1997)\markcite{ww97} (cf.
Kylafis \& Shapiro 1983\markcite{ks83}).
Simplifications that are standard result from
recognizing that the spectral line breadth is much greater than the Zeeman
splitting for SiO masers.

To understand the potential magnitude of the circular polarization that can
be created, first consider the simplified example in which the direction of
the magnetic field rotates uniformly along the path of a ray in an 
unsaturated maser.  Because the weak circular polarization can be treated 
as a perturbation, the analytic solutions for the
linear polarization in Wallin \& Watson (1997)\markcite{ww97} can be 
utilized.  In the unsaturated limit for a fractional linear polarization 
$m$ less than about 1/2, the fractional circular polarization
reduces to
$V/I=(\theta_i/\theta_r)m^2/4$ when the magnetic field rotates by
about one radian along the ray. Here, the $\theta$'s are the
imaginary and real parts of the line shape function which are
similar in magnitude at representative locations within the
spectral line except near line center where $\theta_i=0$. For the
typical $m=1/3$, the resulting $V/I=0.03$ is in excellent agreement
with the average $V/I$ that is observed.

To provide a plausible quantitative description for the variations of
the magnetic fields as a basis for assessing in more detail how well these
can lead to the observed magnitudes and patterns of the polarizations, we
adopt a Kolmogorov-like dependence for the power spectrum of the magnetic
field. Kolmogorov spectra tend to appear in various contexts, including
simulations for other astrophysical environments (e.g., Stone et
al. 1996). Specifically, we utilize the form
$<\mid$\boldmath$A$\unboldmath\hspace{2pt}$_k\mid^2>=C(k^2+k^2_{\rm
min})^{-17/6-b_{\rm e}}$ for the power spectrum of the vector
potential as a function of wave number $k$. Here $b_{\rm e}=0$ corresponds 
to
the strict Kolmogorov variation. The constant $C$ is chosen to yield the 
desired
rms value for the field. Representative field configurations (or
``realizations'') are constructed by statistical sampling for these
Fourier components from a Gaussian distribution. A turbulent velocity 
field, which would likely accompany
a turbulent magnetic field, is created in the same manner. These two fields
are created independently. These
methods are standard and are described in detail elsewhere
(e.g., Wallin, Watson, \& Wyld 1998)\markcite{www98}.
The magnetic and velocity fields are obtained in this manner at
the grid points of a $128\times 128\times 256$ rectangular array. The 
coupled differential equations of radiative transport for the Stokes
intensities are
then integrated numerically for rays that propagate parallel to the  axis
with 256 grid points to give the Stokes intensities for the
$128\times 128$ rays
that emerge perpendicular to that face of the rectangular volume. Although
radiative saturation is not required for creating circular polarization by
the mechanism under investigation, we do attempt to include the effects of
saturation in at least an approximate way for the minority of rays for 
which
saturation is significant. We treat each ray as if it is an independent
``linear maser''. That is, the radiative intensities used for computing the
molecular populations that are needed for the transfer equations of a
particular ray are those of the ray itself. Treating saturation does tend 
to increase somewhat the fractional polarizations.

Within the simplifications of this calculation, the results do not
depend upon the actual magnitude of the magnetic field, only upon its 
direction.
Four quantities  must be specified to perform the calculations -- the slope
of the power spectrum of the fields as expressed by $b_{\rm e}$, the ratio of the 
rms
turbulent velocities to the rms thermal velocity of the SiO molecules, the
pumping rate averaged over all magnetic substates, and the difference
between this pump rate for the $m=\pm 1$ and for the $m=0$ magnetic
substates (the anisotropy). A power spectrum ($b_{\rm e}=1/3$) that is 
steeper than
Kolmogorov seems to yield a somewhat more regular pattern
for the map of the linear polarization. We present computations
for a ratio of four for the rms turbulent to thermal velocity.
This turbulence is modest and subsonic since the sound speed is
determined mainly by the hydrogen. Including turbulent velocities does not 
alter
significantly the average fractional circular polarization that
can be created, but it does provide the opportunity to explore the
appearance of the maps and of the spectral line profiles. The
substate averaged pump rate is reflected in the intensity of the
maser radiation. Our choice is determined by the requirement that
the brightness temperature of the stronger rays be approximately
what is typical of the observations (about $10^{10}$ K). The
anisotropy in the pumping is adjusted so that the computations
yield the average linear polarization of one-third that is
observed. The simplification is made that the pumping is
independent of velocity and of the orientation of the magnetic
field. To be compatible with the observational selection based on
intensity, we consider approximately the strongest 300 rays for
creating our maps and obtaining averaged quantities.

Maps of the linear polarization that we calculate based on a
representative statistical realization are presented in Figures~\ref{fig1}a
and~\ref{fig1}b. Another simplification in our calculations is the omission
of the velocity gradients that would likely be associated with an
outflowing wind in these stellar envelopes and would cause the ring 
structure that is evident in the
analogous observational maps (e.g., KD). Our maps can
be viewed as more indicative of a portion (perhaps one-quarter) of
the observed ring. The tendency in the observations for the pattern of the
linear polarization to vary smoothly despite the turbulence in the 
magnetic field appears in the simulations, as does
the observed occurence of occasional rapid changes in its
direction between nearby locations. The addition of a uniform
magnetic field in Figure~\ref{fig1}b does, as expected, increase the
tendency for order in the pattern of the polarization. The tendency for a
relatively smooth pattern in the linear polarization despite the
turbulence can be understood as a result of the Kolmogorov-like
form in which the variations with the longest wavelengths (here,
the dimensions of the rectangular volume) have the largest amplitudes. The
``richness'' in the possible variations of the polarization
characteristics across a spectral line are evident in  Figures~\ref{fig2}
and~\ref{fig3}. Ordinarily the profile of the linearly polarized radiation
is similar to that for the total intensity, though the position
angle can change by tens of degrees and sometimes by much more
across the spectral line. In contrast, the profile of the Stokes
circular polarization V can either be mostly of one sign and
similar in shape to that of the intensity, or it can be an antisymmetric
``S-curve'' ordinarily associated with the Zeeman effect.
If there were no velocity gradients in our calculations, the
V-spectra would always have the shape of the S-curve even though
the circular polarization is not due to the Zeeman effect. Note
that the computed V-spectra are often shifted somewhat relative to
the profiles for the intensity. The main result here is, of course,
that fractional circular polarization is calculated without the
Zeeman effect and is comparable to what is observed for plausible
assumptions about the circumstellar environment. In agreement with
observations, the fractional circular polarization averaged over the
maps is about three percent, though it can range up to much larger
values at certain locations -- up to more than twenty percent
within the spectral line as shown in Figure~\ref{fig2}. The requirement for
the calculation here is that $g\Omega$ be somewhat greater than the
larger of $R$ and $\Gamma$ (which are comparable for the computations  in
the Figures). Making the usual assumption that $\Gamma$ is
approximately the inverse radiative lifetime (5~s$^{-1}$) for a
vibrational transition of the SiO molecule then leads to the
requirement that the magnetic field be greater than only  about 10--30~mG 
for the observed circular polarization, instead of being equal to the
10--100~G that would be implied by the
Zeeman interpretation. If the magnetic field were weaker than
about 10~mG, the optical axes of the medium would not be determined
by the magnetic field and their orientations would thus not follow
the changes in the directions of the field. Since circular polarization is
created from the linearly polarized radiation (and, specifically $V/I\propto
m^2$ in the simplified example in the foregoing), a close correlation might
be expected between the linear and circular polarization in the radiation of
individual masing features. This is not found to occur in our calculations. The
statistical variations in the velocity and in the magnetic field apparently
introduce sufficient scatter that this correlation is effectively destroyed.
For the maps as a whole, however, the average fractional circular polarization
does tend to increase with the average fractional linear polarization. If
circumstellar (or other) environments with a large number of masing features
occur for which the average of $V/I$ is much greater than the average of
$m^2/4$, it seems most likely that the circular polarization of these would be
due to causes other than those described in this Letter. It would probably be
due to the standard Zeeman effect. Discriminating between the Zeeman and this
non-Zeeman effect based on the observed spectral line profiles of the
polarized radiation does not seem likely since the intrinsic profiles
for Stokes V due to the two effects are similar (antisymmetric ``S-curves'').
These intrinsic profiles can be altered in the emergent radiation for both
as a result of saturation together with gradients in the velocity and the
magnetic field along the line of sight. Finally, we cannot 
exclude the possibility that non-Zeeman effects similar to those in this
Letter can contribute to the observed 
circular polarization of the 22 GHz water masers.

\acknowledgments

This research has been supported in part by NSF Grant AST94-01348. We
are grateful to A. J. Kemball, A. B. Menshchikov and H. W. Wyld for helpful
discussions and information.

\clearpage

\figcaption[wwfig1.eps]{
(a) -- Map for the direction and magnitude of the linearly
polarized fraction of maser radiation calculated for a medium with
turbulent velocity and magnetic fields as described in the text.
The longest vector corresponds to a fractional polarization (averaged
over the spectral line) of 72\%; the average is 32\% for all locations
that are shown. This map is intended to represent a portion (perhaps
one-quarter) of a circumstellar shell  as shown, e.g., in Figure 2 of KD. 
The
largest fractional circular polarization is 18\% and the
average is 3.2\%. (b) -- Same as (a) except that
here a constant magnetic field is present in addition to the
turbulent magnetic field. The constant field is parallel to the
plane of the map and has a magnitude that is twice the rms value
of the turbulent magnetic field.
\label{fig1}}

\figcaption[wwfig2.eps]{
Spectra as a function of Doppler velocity for
selected rays showing the normalized intensity $I$, the fractional
circular polarization $V/I_{\rm p}$ where $I_{\rm p}$ is the peak
intensity within the spectral line, the fractional linear
polarization $m_{\rm p}=(Q^2+U^2)^{1/2}/I_{\rm p}$ [for Stokes $Q$ and 
$U$], and the
variation of the position angle $\chi$ for the linear polarization.
The Doppler velocity is in units of ($2k_{\rm b}T/M$) where $T$ is
the kinetic temperature, $M$ is the mass of an SiO molecule, and
$k_{\rm b}$ is Boltzmann's constant. These spectra are shown for
the rays in Figure 1a with the largest peak intensity (upper), the
largest fractional linear polarization (middle), and the largest
fractional circular polarization (lower).
\label{fig2}}

\figcaption[wwfig3.eps]{
Same meaning as Figure 2 except that here the spectra are shown
for four thick rays in the enlarged area in Figure 1a for which the
directions of the linear polarization are different.
\label{fig3}}
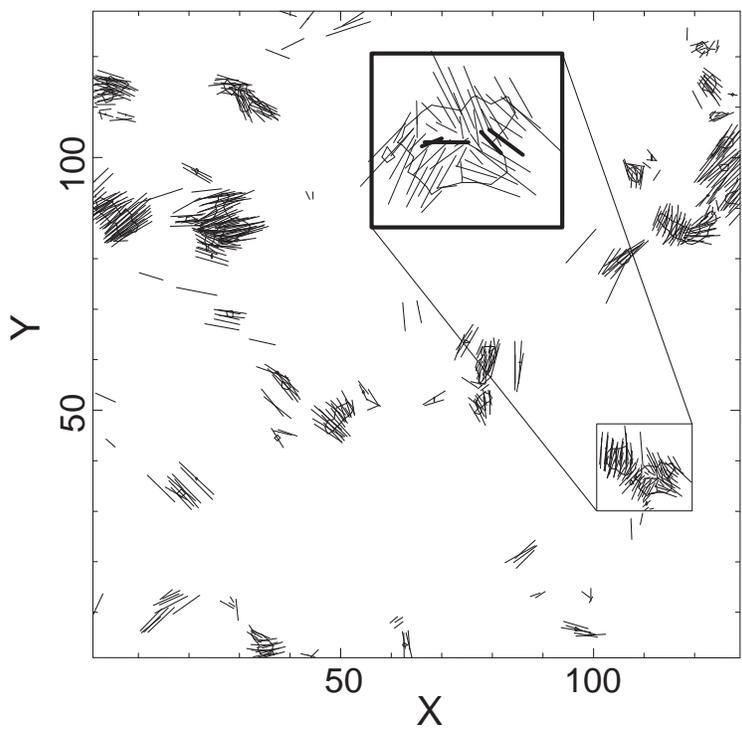
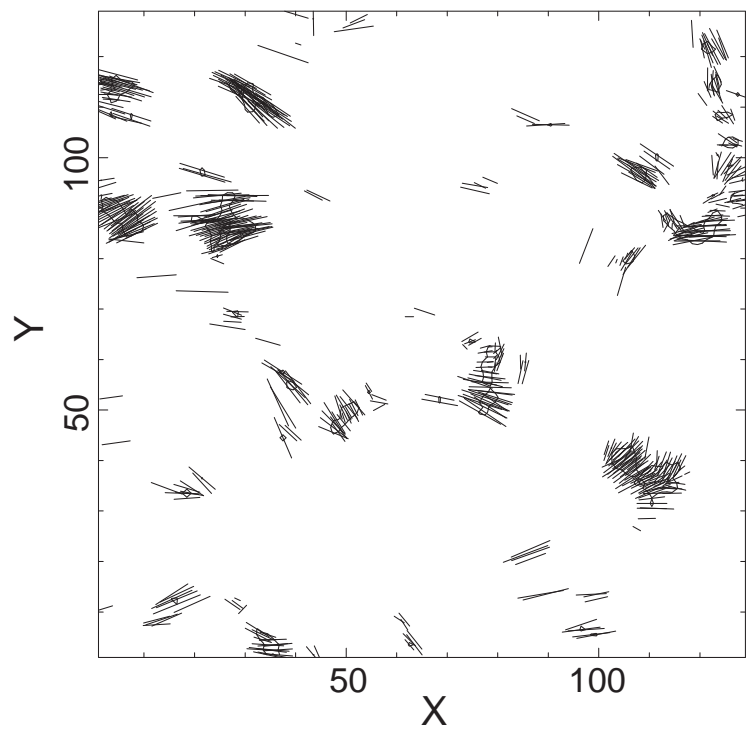
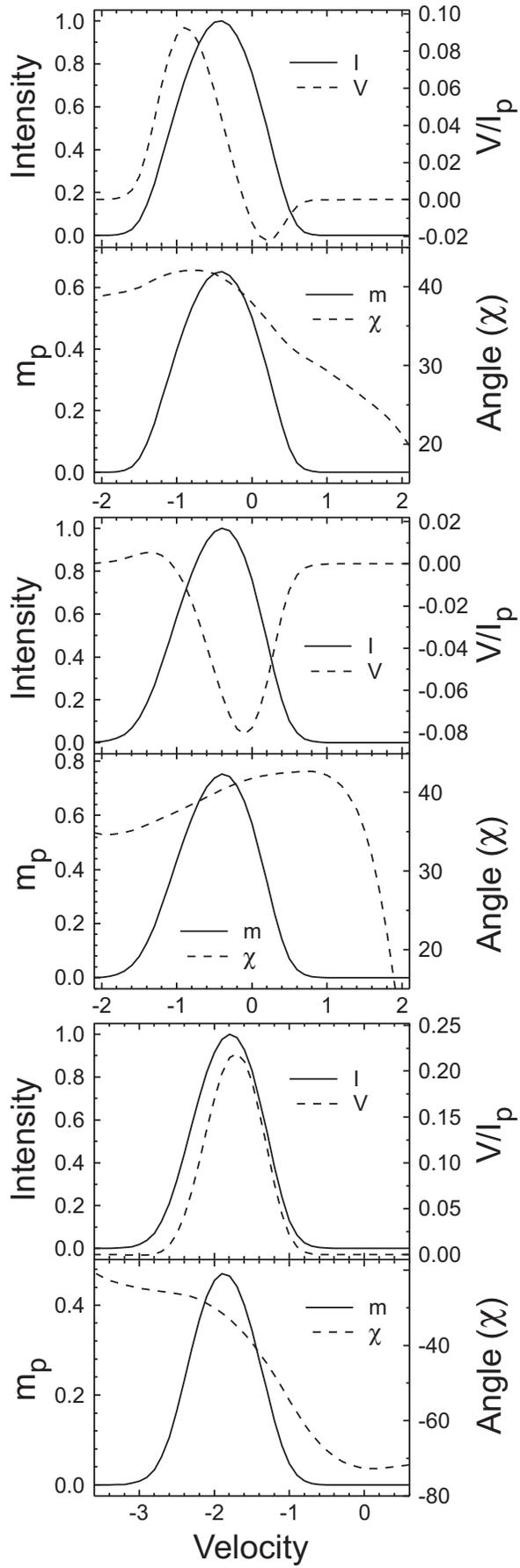
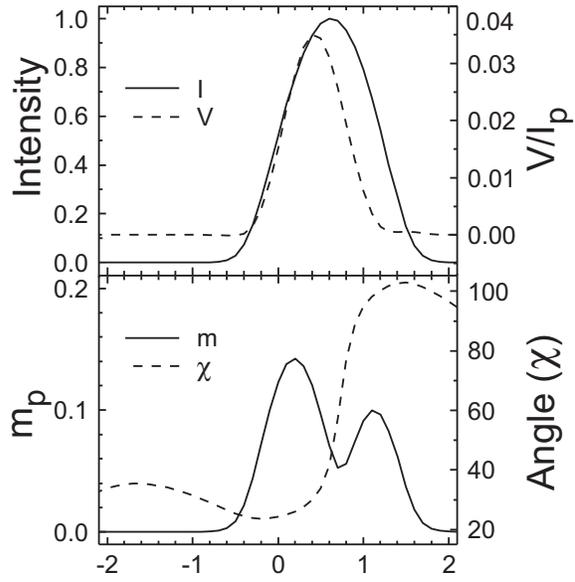
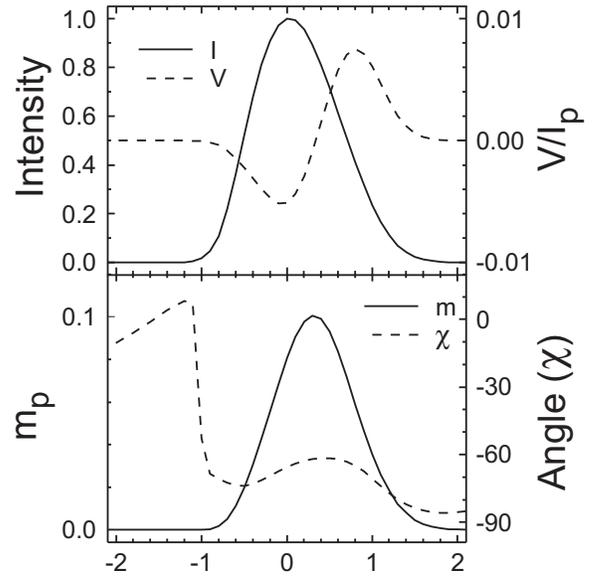
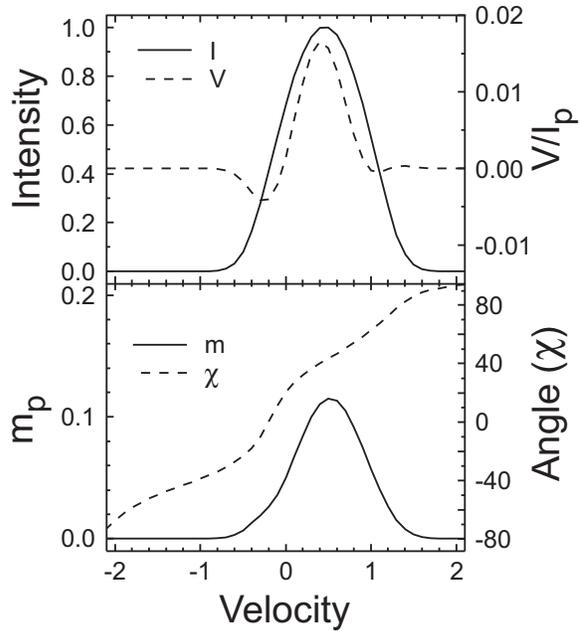
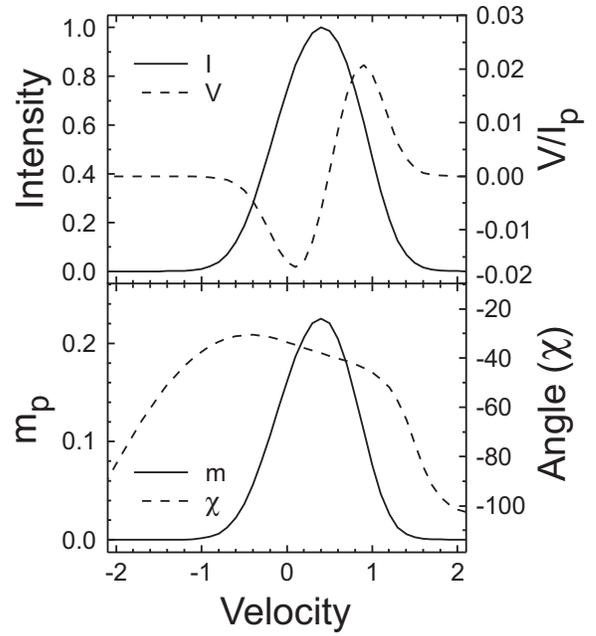
\end{document}